\documentclass[10pt, twocolumn]{article}
\usepackage{wrapfig}
\usepackage{graphicx}

\title{Internet of Things: Concept, Building blocks, Applications and Challenges}
\author{Riad Abdmeziem, Djamel Tandjaoui}
\begin{document}
\maketitle
\begin{abstract}
Internet of things (IoT) constitutes one of the most important technology that has the potential to affect deeply our way of life, after mobile phones and Internet. The basic idea is that every objet that is around us will be part of the network (Internet), interacting to reach a common goal. In another word, the Internet of Things concept aims to link the physical world to the digital one. Technology advances along with popular demand will foster the wide spread deployement of IoT's services, it would radically transform our corporations, communities, and personal spheres. In this survey, we aim to provide the reader with a broad overview of the Internet of things concept, its building blocks, its applications along with its challenges.   
\end{abstract}
\section{Introduction}
\paragraph{}
During the past few years, the acess to the Internet has evolved from static (desktop) access to a more mobile and dynamic one, using several devices - such as Mobile phones,Tablets, Televisions, etc. In this context a novel paradigm named \emph{Internet Of Things (IoT)} is rapidly gaining ground. The basic idea is that every objet that is around us will be part of the network (Internet), interacting to reach a common goal, in another word, the Internet Of thing concept aim to link the physical world to the digital one.
\paragraph{}
The pervasive presence around us of various wireless technologies - such as Radio-Frequency IDentification (RFID)
tags, sensors, actuators, mobile phones, etc. – in which computing and communication systems are seamlessly embedded will form the building block of the IoT concept \cite{ref1}. The IoT's full deployement will give rise to new opportunities for the Information and Communication Technologies (ICT) sector, paving the way to new applications, providing new ways of  working; new ways of interacting; new ways of entertainment; new ways of living \cite{ref2}.
\paragraph{}
Technology advances along with popular demand will foster the wide spread deployement of IoT's services, it would radically transform our corporations, communities, and personal spheres. From the perspective of a private user, IoT's introduction will play a leading role in several services in both working and domestic fields -such as Domotics, e-health, e-learning, security and surveillance, etc. In the same manner, from the business point of view, IoT will bring a deep change in the way automation, industrial manufacturing, logistics, business/process management and transportation of goods and people are handleled.
\paragraph{}
Implementation of IoT paradigm rely on the integration of RFID systems, Wireless Sensor Networks, intelligent technologies (using knowledge to solve certain problems and mainly covering Artificial Intelligence \cite{ref3}) and nanometer technologie (concentrating on the characteristic and application of materials of size between 0.1 and 100 nm). Up to this day, 
since research of IoT is still embryonic there exist no common IoT architecture.
Nowadays the Electronic Product Code (EPC) network architecture supported by EPCglobal \cite{ref4} together with the Unique/Universal/Ubiquitous IDentifier (UID) architecture in Japan \cite{ref5} are the most representative among others. The main idea underlying EPCglobal network is to use RFID and wireless technologies to wrap every day's live objects and connect them to the \emph{traditional} Internet, while, UID provides middleware based solutions for a global visibility of objects.
\paragraph{}
Several challenges stand between the conceptual idea of IoT and its full deployement into our daily life. Main issues are : making complete interoperability of heteregenous interconnected devices which require adaptation and autonoumous behaviour while guaranteeing trust, privacy, and security; networking aspect is not in rest, low computation and energy capacities that characterized \emph{the things} of the IoT bring ressource effeciency as a fundamental element in the proposed solutions. Around the globe, several industrial, standardization and research bodies are currently involved in the devloppement of solutions in order to bring answers to the  highlighted technological requirements.
\paragraph{}
In this survey, we aim to provide to the reader an overview of the IoT concept, the different enabling technologies, research challenges and the implications of a wide spread diffusion of IoT.
The remainder of this paper is organized as follows: in Section 2, definitions of IoT from various perspectives are introduced. Section 3 introduces the main IoT enabling technologies. The applications of IoT already available are summarized in Section 4. Section 5 states the research challenges. Finally Section 6 gives the conclusion.
\section{Definition and vision}
In Internet of things (IoT), huge number of small devices will be connected to the Internet in some way. IoT's definition is usually studied through various perspectives. According to \cite{ref3}, the IoT paradigm shall be the result of the convergence of three main visions: internet-oriented (middleware), things oriented (sensors) and semantic-oriented (knowledge) as shown in Figure 1.
\begin{figure}
\begin{center}
\includegraphics[scale = 0.4]{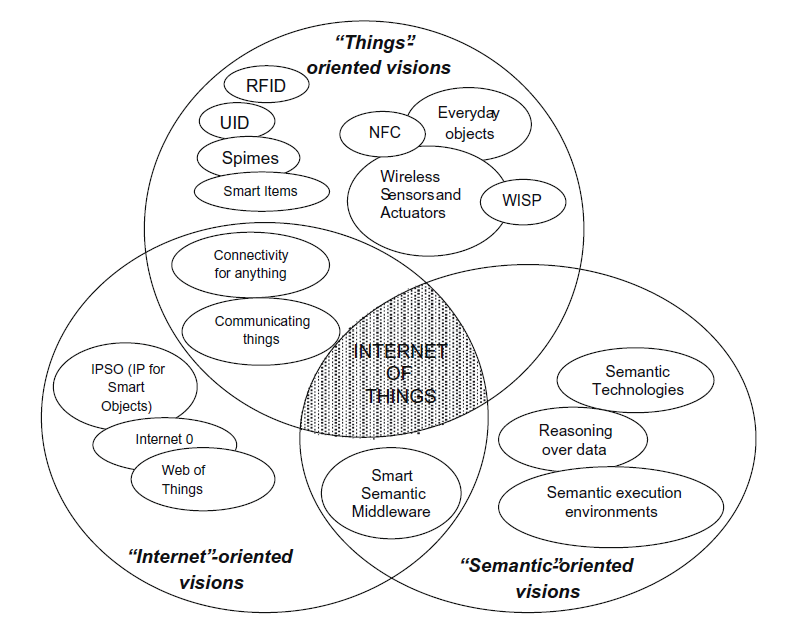}
\end{center}
\caption{Internet of Things paradigm as a result of the convergence of different visions}
\end{figure}
\emph{Perspective of Things: } This perspective focuses on how to integrate generic "objects" or "Things" into a common framework, and the "Things" under investigation are RFID tags. RFID is considered as a one of the leading technologies \cite{ref6}, mainly due to its maturity and low cost, and consequently its strong support from the business community. Nevertheless, IoT is more than a global EPC system where the only objects are RFID tags. Besides that,United Nations (UN) has also stated that the perspective of "Things" of IoT goes beyond RFID. It is stated in a UN report that a new era of ubiquity is coming where the users of the Internet will be counted in billions, and where humans may become the minority as generators and receivers of traffic. Changes brought about by Internet will be dwarfed by those prompted by the networking of everyday objects.
\paragraph{}
The consortium CASAGRAS \footnote{Coordination And Support Action for Global RFID-related Activities and Standardisation} also proposes an IoT vision statement that goes well beyond a mere "RFID centric" approach. Its members focus on a world where things can automatically communicate to computers and each other, providing services to the benefit of the human kind. It not only proposes IoT would connect both virtual and physical generic objects as a global infrastructure, but also emphasizes the importance of incorporating the traditional Internet related technologies and infrastructures in the development of IoT.
Similarly, other relevant institutions have stressed the concept that IoT has primarily focused on the "Things" and that
the road to its full deployment has to start from the augmentation in the Things’ intelligence.
From the thing's perspective, the International Telecommunication Union (ITU) has given the following definition of the IoT: from anytime, any place connectivity for anyone, we will add a connectivity for anything . The same definition is given by the European Commission, it relates to : things having identities and virtual personalities operating in smart spaces using intelligent interfaces to connect and communicate within social, environmental, and user contexts \cite{ref7}.
\paragraph{}
\emph{Perspective of Internet: }
``A world where things can automatically communicate to computers and each other providing services to the benefit of the human kind'', this vision is brought by the CASAGRAS consortium, a vision of IoT as a global infrastructure which connects both virtual and physical generic objects and highlights the importance of including existing and evolving Internet and network developments in it. In this sense, IoT becomes the natural enabling architecture for the deployment of independent federated services and applications, characterized by a high degree of autonomous data capture, event transfer, network connectivity and interoperability. This definition leads to the Internet oriented vision of IoT, while the perspective of things focuses on integrating generic objects into a commun framework, the perspective of "Internet" pushes towards a network oriented definition. 
\paragraph{}
Within the latter category falls the IoT vision of the IPSO(IP for Smart Objects) Alliance \cite{ref8}, a forum formed in September 2008 by 25 founding companies to promote the Internet Protocol as the network technology for connecting Smart Objects around the world.
This vision favors the Internet protocols as the network technology for connecting smart objects around the world.
According to IPSO, the IP stack is a lightweight protocol that already connects a huge amount of communicating devices and runs on tiny and battery operated embedded devices. Reducing the complexity of the IP stack in order to design a protocol to route \emph{IP over things} is the promoted idea.
\paragraph{}
Furthermore, as the Internet is running out of addresses, in the near future it will be moving to a new protocol, IPv6. The current system, IPv4, has roughly four billion addresses. The new address space can support (about 3.4×1038) addresses, which means, to take a commonly used analogy, that it provides enough addresses for every grain of sand on every beach in the world! While it is unlikely that we will be assigning IP addresses to grains of sand, the idea of assigning them to each of the more or less 5,000 daily objects that surround us, is quite appealing. With the right technology in each object (e.g., an RFID tag) and the right network in the surroundings, it will become easy to locate and catalogue items in a few seconds and to reap the benefits of the vast array of new information that communications among them will provide. IPv6 is undoubtedly one of the steps to making the Internet of Things a reality\cite{ref15}. 
\paragraph{}
Reducing the complexity of the IP stack and incorporating IEEE 802.15.4 into the IP architecture is looked as the wisest way to to move from the Internet of Devices to the Internet of Things. According to the IPSO and 6LoWPAN \footnote{IPV6 Low Power Wireless Personnal Area Network}, the IoT will be deployed by means of a sort of simplification of the current IP to adapt it to any object and make those objects addressable and reachable from any location.
\paragraph{}
\emph{Perspective of Semantics: } The basic idea behind is that the number of items involved in the futur Internet Of Things is likely to become very high, thus, issues related to how to represent, store, interconnect, search, and organize information generated by the IoT will become very challenging.
\paragraph{}
Such development will also necessarily create demand for a much wider integration with various external resources,
such as data storages, information services, and algorithms, which can be found in other units of the same organization,
in other organizations, or on the Internet. Therefore, issues of representing, storing, searching, and interconnecting information generated in IoT will become very challenging.

In this context, semantic technologies could play a key role. In fact, these can exploit appropriate modeling solutions for things description, reasoning over data generated by IoT, semantic execution environments and architectures that accommodate IoT requirements \cite{ref1}.
\section{IoT elements}

The Internet of things is unlikely to rise as a brand new class of systems. An incremental developement path, along which IoT technologies will be progressivly employed to extend existing ICT systems/applications, providing additionnal functionalities related to the ability of interacting with the physical realm.
This section focuses on the enabling technologies that are expected to form the building blocks of the IoT, each technology is briefly presented with its  supposed impact on the IoT's devloppement. Figure 2 gives an overview of the main technologies that will be involved in the future IoT.
\begin{figure}
\begin{center}
\includegraphics[scale = 0.4]{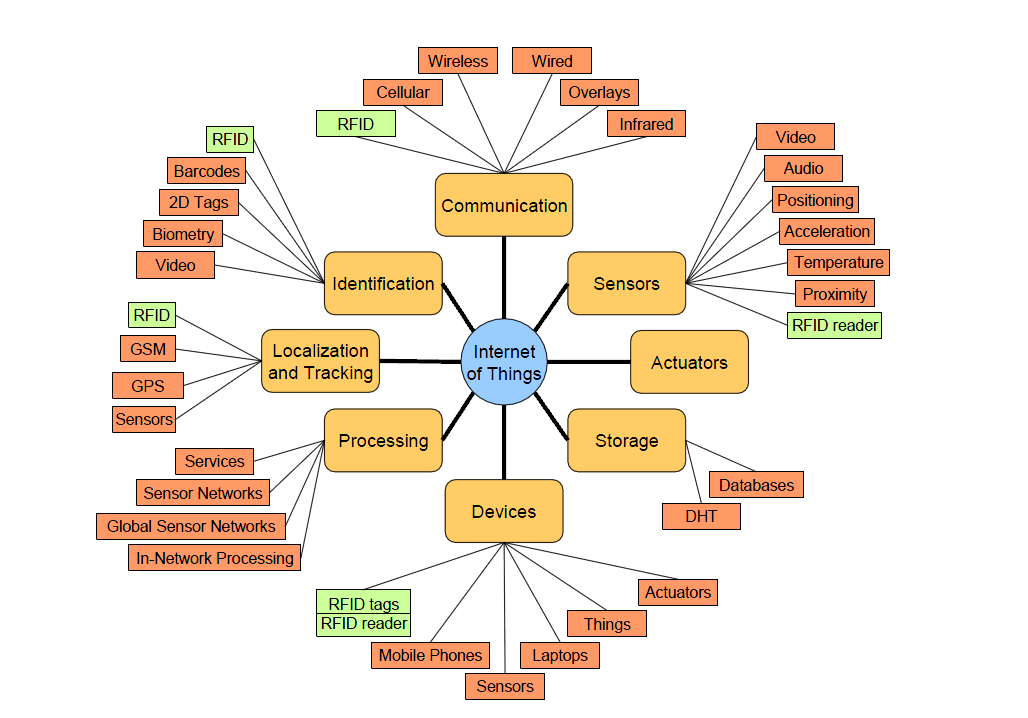}
\end{center}
\caption{IoT Elements}
\end{figure}
\subsection{Sensing, computing and identification technologies} 
The ability of sensing the environement and to self-organize into ad hoc networks represent an important feature from the IoT perspective. Nevertheless, some limiting factors stands in front of a widespread adoption:
Energy management of the futur embedded devices is a crucial issue, in order to get minimal computationnal capablilties sensor nodes will have to be equipped with a battery. While a number of solutions for increasing energy efficiency – at various layers of the OSI model – has been devised, the need to replace batteries from time to time represents a huge barrier to the widespread development of IoT. Besides, nodes in a classical wireless sensor network are expected to possess a set of common characteristics, and to share a number of common features including a full protocol stack. While advances in embedded electronics and software are making such a requirement less and less stringent, solutions able to accommodate heterogeneity in terms of supported features should be introduced to ease incremental deployment \cite{ref2}.
\paragraph{}
Wireless technologies will play a key role in the futur IoT, in a way where the major part of data traffic between objects will be carried in a wireless way. Otherwise the reduction in terms of size, weight, energy consumption, and cost of the radio can lead us to a new era where radios could be integrated in almost all objects and thus, add the world ‘‘anything” to the ‘‘Anytime, anywhere, anymedia” vision.
\paragraph{}
Wireless Sensor Networks (WSN) and radio-frequency identification (RFID) are considered as the two building blocks of the sensing and communication technologies in the futur IoT:
\subsubsection{RFID}
RFID technology is a major breakthrough in the embedded communication paradigm which enables design of microchips for wireless data communication. RFID tags are expected to play a key role as enabling identification technology in IoT. They help in automatic identification of anything they are attached to acting as an electronic barcode.
From a physical point of view, as shown in Figure 3 ,a RFID tag is a small microchip attached to an antenna (that is used for both receiving the reader signal and transmitting the tag ID) in a package which usually is similar to an adhesive sticker. Dimensions can be very low: Hitachi has developed a tag with dimensions 0.4 mm x 0.4 mm x 0.15 mm.
\begin{figure}
\begin{center}
\includegraphics[scale = 0.3]{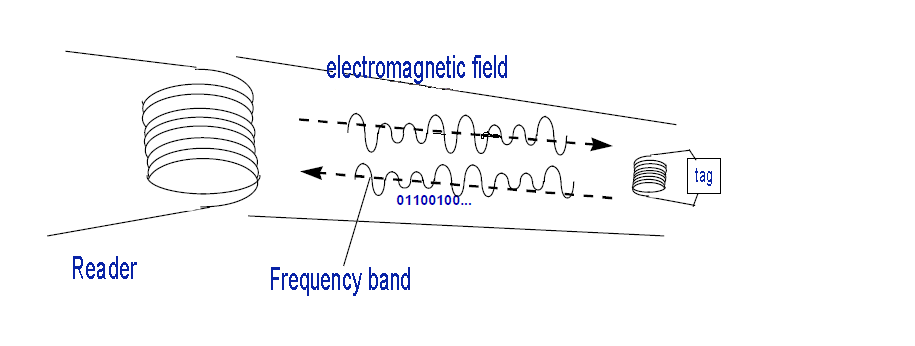}
\end{center}
\caption{RFID tag and reader}
\end{figure}
\paragraph{}
The passive RFID tags are not battery powered and they use the power of the reader‘s interrogation signal to communicate the ID to the RFID reader. This has resulted in many applications particularly in retail and supply chain management. The applications can be found in transportation (replacement of tickets, registration stickers) and access control applications as well. The passive tags are currently being used in many bank cards and road toll tags which is among the first global deployments. Active RFID readers have their own battery supply and can instantiate the communication. Obviously the radio coverage is the highest for active tags, though, this is achieved at the expenses of higher production costs . Of the several applications, the main application of active RFID tags is in port containers for monitoring cargo \cite{ref1}.

Sensor networks will also play a crucial role in the IoT. In fact, they can cooperate with RFID systems to better track the status of things, i.e., their location, temperature, movements, etc. As such, they can augment the awareness of a certain environment and, thus, act as a further bridge between physical and digital world.
\subsubsection{WSN}
Sensor networks consist of a certain number, which can be very high, of sensing nodes communicating in a wireless multi-hop fashion \cite{ref16} as shown in Figure 4.
\begin{figure}
\begin{center}
\includegraphics[scale = 0.2]{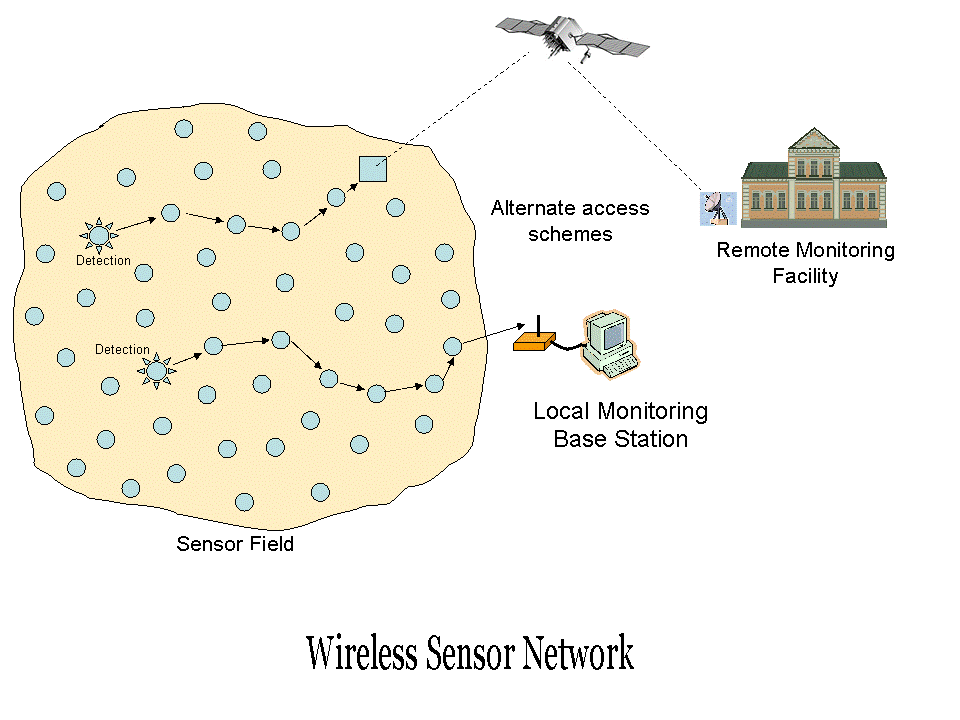}
\end{center}
\caption{Wireless Sensor Network}
\end{figure}
Usually nodes report the results of their sensing to a small number (in most cases, only one) of special nodes called sinks. A large scientific literature has been produced on sensor networks in the recent past, addressing several problems at all layers of the protocol stack. Design objectives of the proposed solutions are energy efficiency (which is the scarcest resource in most of the scenarios involving sensor networks), scalability (the number of nodes can be very high), reliability (the network may be used to report urgent alarm events), and robustness (sensor nodes are likely to be subject to failures for several reasons). Today, most of commercial wireless sensor network solutions are based on the IEEE 802.15.4 standard, which defines the physical and MAC layers for low-power, low bit rate communications in wireless personal area networks (WPAN). IEEE 802.15.4 does not include specifications on the higher layers of the protocol stack, which is necessary for the seamless integration of sensor nodes into the Internet.
\paragraph{}
Integration of sensing technologies into passive RFID tags would enable a lot of completely new applications into the IoT context. Sensing RFID systems will allow to build RFID sensor networks, which consist of small, RFID-based sensing and computing devices, and RFID readers, which are the sinks of the data generated by the sensing RFID tags and provide the power for the network operation.

\subsection{Middleware}
The middleware is a software layer or a set of sub-layers interposed between the technological and the application level. Its main feature of hiding the details of different technologies is fundamental to keep the programmer away from issues that are not directly pertinent to her/his focus, which is the development of the specific application enabled by the IoT infrastructures. The middleware is gaining more and more importance in the last years due to its major role in simplifying the development of new services and the integration of legacy technologies into new ones. This spares the programmer from the exact knowledge of the heterogeneous set of technologies adopted by the lower layers\cite{ref3}.
\paragraph{}
As far as frameworks for developing IoT applications are concerned, a major role is expected to be played by approaches
based on service-oriented computing(SOC). SOC proposes a possibly distributed architecture, whereby entities are treated in a uniform way and accessed via standard interfaces providing a common set of services and an environment for service composition. Figure 5 addresses the middleware issues with a complete and integrated architectural approach. Each layer is briefly presented:
\begin{figure}
\begin{center}
\includegraphics[scale = 0.5]{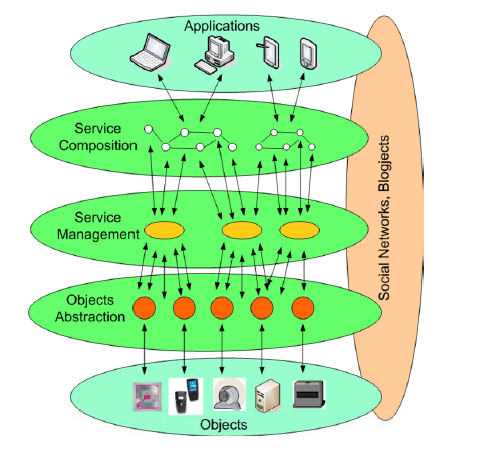}
\end{center}
\caption{IoT middleware architecture}
\end{figure}
\paragraph{}
\emph{Applications: } Applications are on the top of the architecture, exporting all the system’s functionalities to the final user. the integration between distributed systems and applications is ensured through the use of standard web service protocols and service composition technologies.\\

\emph{Service composition: }This is a common layer on top of a SOA-based middleware architecture. It provides the functionalities for the composition of single services offered by networked objects to build specific applications. On this layer there is no notion of devices and the only visible assets are services.\\

\emph{Service management: }This layer provides the main functions that are expected to be available for each object and that allow for their management in the IoT scenario. This layer might enable the remote deployment of new services during run-time, in order to satisfy application needs. A service repository is built at this layer so as to know which is the catalogue of services that are associated to each object in the network.\\

\emph{Object abstraction: } The IoT relies on a vast and heterogeneous set of objects, each one providing specific functions accessible through its own dialect. There is thus the need for an abstraction layer capable of harmonizing the access to the different devices with a common language and procedure.

A service-oriented architecture (SOA) is mainly a collection of services, which communicate with each other via a set of standardized interaction patterns. The communication can involve either simple message passing or it could involve two or more services coordinating some activity via appropriate protocols. Currently, many SOC deployments make use of Web-based protocols (e.g., http) for supporting interoperability across administrative domains and enabling technologies. SOC can be used to manage web services and make them act like a virtual network, adapting applications to the specific users needs.
Service-oriented architectures support a given level of heterogeneity and flexibility in the software modules, nevertheless, to be deployed and executed, SOC/SOA in general and Web services in particular, cannot be straightforwardly applied to the construction of IoT applications. In particular, such approaches, at least in their current form, may prove too heavyweight for being deployed on resources-constrained devices.
Nonetheless, they represent a very powerful approach in terms of abstracting functionality from the specific  software implementation as well as for ensuring integration and compatibility of IoT technologies into the bigger Future Internet-Future Web perspective\cite{ref2}.
\subsection{Ambient intelligence and self management systems} 
A parralel can be established between IoT and ambient intelligence, as a matter of fact, IoT shares a number of characteristics with ambient intelligence. In Ambient Intelligence (AmI), environments rich in sensing/computing/actuation capabilities are designed so to respond in an intelligent way to the presence of users, thereby supporting them in carrying out specific tasks. Ambient intelligence builds upon the ubiquitous computing concept, loosely defined as the embedding of computational devices into the environment \cite{ref9}.
\paragraph{}
AmI shares with IoT a number of aspects. This comprises the inclusion in the system of sensing/computing capabilities embedded in the environment. Nevertheless, AmI applications have been mainly developed for closed environments (e.g., a room, a building), whereby a number of specific functions, known at design time, can be accommodated and supported. Accordingly, one of
the main focus of research in AmI has been the development of reasoning techniques for inferring activities of users and proposing appropriate response strategies from the embedded devices. IoT expands the AmI concepts to integrate open scenarios, whereby new functions/capabilities/ services need to be accommodated at run-time without them having been necessarily considered at design time. This requires IoT solutions to be inherently autonomic, i.e., presenting the self-configuration and self-organization, possibly cognitive, capabilities needed to provide this additional degree of flexibility.
\paragraph{}
IoT application scenarios require applications to prove adaptable to highly diverse contexts, with different resources available and possibly deployment environments changing over time. A number of approaches have been proposed to overcome devices heterogeneity in related scenarios. All the efforts required in terms of development of IoT architectures, methods for management of resources, distributed communication and computation, represent the baseline for the introduction of innovative services that will improve user's experience.
\paragraph{}
IoT services will be responsive in nature, being able to anticipate user needs, according to the situation they are in, by means of dynamic resource management schemes and on-the-fly composition of different service components. This requires applications to be able to understand the context and situation the user is in. Such a theme has been addressed within the ambient intelligence, ambient assisted living and pervasive computing fields, leading to a number of solutions able to leverage contextual information coming from a number of sources. 
\paragraph{}
Services in IoT are expected to be able to seamlessly adapt to different situations and contexts. A number of research efforts for building self-adaptive situated services have been undertaken in the last few years \cite{ref10}. However, we are still far from reaching a global understanding of how to develop self-adaptive services presenting the flexibility level required by IoT scenarios. Most of the approaches proposed have been conceived to be applied to a single, well-defined specific application field. What is needed to foster the deployment of IoT applications is instead a set of design patterns that can be used to augment end-user applications with self-adaptive properties. This requires methods for discovering, deploying and composing services at run-time in a distributed fashion, supporting autonomicity within all phases of the service life-cycle. While smart objects may be able to run some limited and lightweight services, one key aspect of IoT is the integration with the Internet infrastructure, i.e., the cloud. This may take the form of appropriate Web-based services and applications, able to leverage data and/or atomic services made available by smart things to provide value-added services to the end user.
\section{Applications}
Enabling the objects in our everyday working or living environment to possibly communicate with each other and elaborate
the information collected will make a lot of applications possible.

\begin{figure}
\begin{center}
\includegraphics[scale = 0.4]{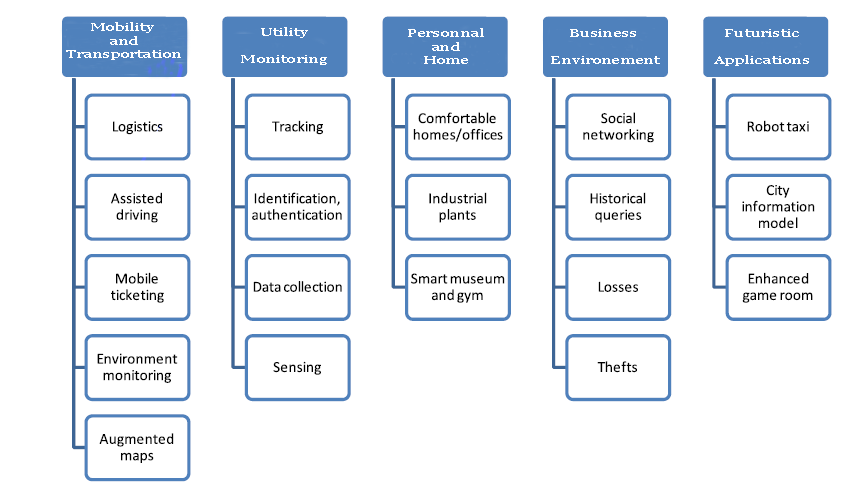}
\end{center}
\caption{Applications domains and relevant major scenarios}
\end{figure}

As shown in figure 6 ,there are several application domains which will be impacted by the emerging Internet of Things. IoT technologies, which are either directly applicable or closer to our current living habitudes, might be classified into the following domains: Personal and social, Business environement, Service and utility monitoring and Mobility and transportation based on the type of network availability, coverage, scale, heterogeneity, repeatability, user involvement and impact.
\subsection{Personal and social} 
In this domain, the sensor information collected is used only by the individuals who directly own the network. Usually WiFi is used as the backbone enabling higher bandwidth data (video) transfer as well as higher sampling rates (Sound).
Ubiquitous healthcare has been envisioned for the past two decades. IoT gives a perfect platform to realize this vision using body area sensors and IoT backend to upload the data to servers. For instance, a Smartphone can be used for communication along with several interfaces like Bluetooth for interfacing sensors measuring physiological parameters. So far, there are several applications available for Apple iOS, Google Android and Windows Phone operating system that measure various parameters.
\paragraph{}
An extension of the personal body area network is creating a home monitoring system for aged-care, which allows the doctor to monitor patients and elderly in their homes thereby reducing hospitalization costs through early intervention and treatment.
Control of home equipment such as air conditioners, refrigerators, washing machines etc., will allow better home and energy management. This will see consumers become involved in IoT revolution in the same manner as the Internet revolution itself.
\paragraph{}
Social networking is set to undergo another transformation with billions of interconnected objects. An interesting development will be using a Twitter like concept where Things in the house can periodically tweet the readings which can be easily followed from anywhere creating a TweetOT. Although this provides a common framework using cloud for information access, a new security paradigm will be required for this to be fully realized \cite{ref1}.
\subsection{Business environement}
We refer to the Network of Things within a work environment as an enterprise based application. Information collected from such networks are used only by the owners and the data may be released selectively. Environmental monitoring is the first common application which is implemented to keep a track of the number of occupants and manage the utilities within the building.
\paragraph{}
Sensors have always been an integral part of factory setup for security, automation, climate control, etc. This will eventually be replaced by wireless system giving the flexibility to make changes to the setup whenever required. This is nothing but an IoT subnet dedicated to factory maintenance.
\paragraph{}
Real-time information processing technology based on RFID and NFC (Near Field Communication) will be widely used in supply chain, due to their low cost and low requirement. Accordingly, accurate and real-time information relating to inventory of finished goods, work-in-progress, and in-transit stages with reliable due dates would be obtained. As a result, the demand forecast would be more accurate and extra buffers would be unnecessary.
\paragraph{}
For example, a manufacturer of soft drinks can identify with the click of a button how many containers of its soda cans are likely to reach their expiration date in the next few days and where they are located at various grocery outlets. Using this information, it might modify its future production and distribution plans, possibly resulting in significant cost savings. As a result of applications, the reaction time of traditional enterprises is 120 days from orders of customers to the supply of commodities while Wal-Mart applying these technologies only needs few days and can basically work with zero safety stock \cite{ref7}.
\subsection{Service and utiliy monitoring}
The information from the networks in this application domain are usually for service optimisation rather than consumer consumption. It is already being used by utility companies (smart meter by electricity supply companies) for resource management in order to optimise cost vs. profit. These are made up of very extensive networks (usually laid out by large organisation on regional and national scale) for monitoring critical utilities and efficient resource management. The backbone network used can vary between cellular, WiFi and satellite communication.
\paragraph{}
Smart grid and smart metering is another potential IoT application which is being implemented around the world. Efficient energy consumption can be achieved by continuously monitoring every electricity point within a house and using this information to modify the way electricity is consumed. This information at the city scale is used for maintaining the load balance within the grid ensuring high quality of service.
Video based IoT which integrates image processing, computer vision and networking frameworks will help develop a new challenging scientific research area at the intersection of video, infrared, microphone and network technologies. Surveillance, the most widely used camera network applications, helps track targets, identify suspicious activities, detect left luggages and monitor unauthorized access. Automatic behavior analysis and event detection (as part of sophisticated video analytics) is in its infancy and breakthroughs are expected in the next decade \cite{ref1}.
\paragraph{}
Disaster alerting and recovery systems could be significantly enhanced. Natural disasters (flood, landslide, forest fire, etc.) and accidental disasters (coal mine accident, etc.) are taking place more and more frequently. Technologies in IoT, such like RFID and WSN could play a crucial role in disaster alerting before it happens, and disaster recovery after it ends. In order to lessen the effects of natural disasters such like flood, landslide or forest fire, it is necessary to anticipate its occurrence and to alert in time. The timely access to relevant information on hazardous environmental conditions gives residents in the nearing area time to apply preparedness procedures, alleviating the damage and reducing the number of casualties derived from the event. WSN enables the acquisition, processing and transmission of environmental data from the location where disasters originate to potentially threatened cities. Then this information could be used for authorities to rapidly assess critical situations and to organize resources. As to accident disaster recovery, for example, after a coal mine accident occurs, instant tracking and positioning of trapped workers using RFID technologies could provide timely rescue and lessen casualties and economic loss to the largest extent. Knowing trapped workers geographic distribution and comparatively accurate position, the rescue action would be more targeting thus is time-efficient \cite{ref7}.
\paragraph{}
Water network monitoring and quality assurance of drinking water is another critical application that is being addressed using IoT. Sensors measuring critical water parameters are installed at important locations in order to ensure high supply quality. This avoids accidental contamination among storm water drains, drinking water and sewage disposal. The same network can be extended to monitor irrigation in agricultural land. The network is also extended for monitoring soil parameters which allows informed decision making about agriculture.
\subsection{Mobility and transportation}
Smart transportation and smart logistics are placed in a separate domain due to the nature of data sharing and backbone implementation required. Urban traffic is the main contributor to traffic noise pollution and a major contributor to urban air quality degradation and greenhouse gas emissions. Traffic congestion directly imposes significant costs on economic and social activities in most cities. Supply chain efficiencies and productivity, including just-in-time operations, are severely impacted by this congestion causing freight delays and delivery schedule failures. Dynamic traffic information will affect freight movement, allow better planning and improved scheduling. \cite{ref1}
\paragraph{}
Cars, buses and taxis as well as roads intersections are becoming more instrumented with sensors, actuators, and processing power. Important information could be collected to realize traffic control and guidance, help in the management of the depots, and provide tourists with appropriate transportation information. One of the successful applications of IoT in transportation is the Traffic Information Grid (TIG) implemented on ShanghaiGrid.
\paragraph{}
TIG shields all the complexities in information collection, storage, aggregation and analysis. It utilizes Grid technology to ingrate traffic information collected by sensors and actuators, share traffic data and traffic resources, provide better traffic services to traffic participators, and help to remove traffic bottlenecks and resolve traffic problems. The TIG portal provides users with various information services and can be accessed by Web browsers, mobile phones, PDAs and other public infrastructure. Services provided in TIG included road status information, least-time travel scheme selection, leastcost travel scheme selection, map operation and information query \cite{ref7}.
\paragraph{}
Apart from the above applications, many others could be described as futuristic since they rely on some (communications, sensing, material and industrial processes) technologies that are still to come or whose implementation is still too complex. The most appealing futuristic applications included robot taxi, city information model and enhanced game room.
\section{Research challenges}
In spite of the partial feasability of the IoT concept thanks to the advances realized in the enabled technologies (seen in section 3), a large effort is still required from the research community in order to guarantee a full, functional and safe deployement of IoT in our every day life. In this section we propose a brief description of the main open issues. 
\subsection{Security management and privacy}
In the Internet of Things, everything real becomes virtual, which means that each person ans thing has a localizable, addressable,  and readable countrepart on the internet. The IoT promises to extend (anywhere, anyhow, anytime) computing to (anything, anyone, any service). IoT is unlikely to fullfill a widespread diffusion until it provides stong security foundations which will prevent the growth of malicious models or at least, mitigate their impact. As shown in Figure 7, there are a number of top level security research challenges, some of them are described in this section:
\begin{figure}
\begin{center}
\includegraphics[scale = 0.5]{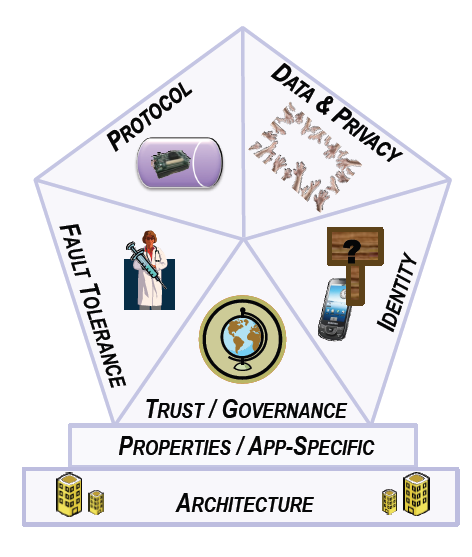}
\end{center}
\caption{Security research challenges in IoT}
\end{figure}
\subsubsection{Data confidentiality} 
Data confidentiality represents a major issue in IoT visions, making sure that only authorized entities  can acces and modify data. This is particularly relevant in the business context, whereby, data may represent a valuable asset that has to be protected. Two important aspects have to be taken in consideration, first the definition of an access control mechansim and second the definition of an object authentication process.
\paragraph{}
As data in IoT applications will be related to the physical realm, ensuring data confidentiality is a main constraint for many uses cases. For instance: a smart community application where a group of homes (located in a close geographic area) are connected and exchange data might rise privacy concerns \cite{ref12}. In such a context, data should be accessible only by the appropriate users (home owners for instance), the leakage of private data into the public sphere comprise in a serious way the user's privacy.
\paragraph{}
Usual solutions for ensuring data confidentiality may not be straightforwardly applied to IoT context due to scalability issues generated by the sheer amount of data in the IoT network. Optimal cryptography algorithms and adequat key management systems, as well as security protocols that connect all the devices  through the Internet form the cornerstone data confidentiality. Although it is possible to implement existing standards (AES for instance), some IoT devices such as passive RFID tags, might be extremly constrained. Cryptography mechanisms must be smaller and faster but with little or no reduction in security level. Mechanisms could include symmetric algorihthms, hash functions and random number generators \cite{ref11}.
\subsubsection{Privacy protection} 
According to \cite{ref2}, privacy defines the rules under which data referring to individual users may be accessed. Privacy is one of the most sensitive subject in any discussion of IoT security. The data availability explosion has fostered some entities to profile and track users without their consent. The IoT's anywhere, anything anytime nature could easily turn such practices into a dystopia. Facebook accounts already affect a user's employability and personal interactions, IoT could certainly raise much more concerns regarding to the huge amount of personal data available.
\paragraph{}
Different approches are in development to protect the personal information of IoT users. The delegation mechanism is one privacy preservation proposal. An unauthorized RFID reader will retrieve only a random value, so it will not be able to track the user.
However, limiting access to the user is not the only protection scenario. In some cases, users will want to provide information without revealing too much about themselves. Some solutions in this context let the user find others who best match his preferences, without actually revealing such preferences \cite{ref11}. For instance, a user can try to locate someone in the neighborhood who like a particular type of music without providing his own location and music preferences.
\subsubsection{Trust and governance}
There is no consensual defintion of the concept of trust, nevertheless, it is a wide used concept in different context related to computer science's security. A widely used definition is the one provided by Blaze and Feigenbaum \cite{ref13}, which refers to security policies regulating accesses to resources and credentials that are required to satisfy such policies.
\paragraph{}
In the IoT context, trust mechansims have to be able to define trust in a dynamic, collaborative environement and provide trust throught an interaction. Another vision of trust encompasses how users feel while interacting in the IoT. Feelings of helplessness and being under some unknown control can greatly hinder the deployement of IoT-based applications and services. Users must be able to control their own services, in another word, there must be support for controlling the state of the virtual world.
\paragraph{}
Governance helps strengthen trust in the IoT. A common framework for security policies will support interoperability and reduce liability. If someone can attribute a malicious action  transaction to a particular user or agent, it will be possible to punish that user or the agent's owner. But governance is a double-edge sword. On the one hand, it offers stability, support for political decisions, and a fair enforcement mechanism. On the other hand, governance can easily become excessive, fostering an 
environment in which people are continuously monitored and controlled. Addressing the challenges of a governance framework when countless stakeholders and billions of objects are connected requires the combined efforts of several research communities\cite{ref11}.

\subsubsection{Fault tolerance}
Clearly, the IoT will be more susceptible to attack than the current Internet, since billions more devices will be producing and consuming services. Highly constrained devices will be the most vulnerable, and malicious entities will seek to control at least some devices either directly or indirectly. In this context, fault tolerance is indispensable to assure service reliability, but any solution must be specialized and lightweight to account for the number of constrained and easily accessible IoT devices. 
Achieving fault tolerance in the IoT will require three cooperative efforts. The first is to make all objects secure by default.
Researchers must work on designing secure protocols and mechanisms, since it might be very costly to provide an update for billions of deployed, heteregenoues and interconnected devices.
\paragraph{}
The second effort is to give all IoT objects the ability to know the state of the network and its services. This system would need to give feedback to many other elements; for example, a watchdog system could acquire data as part of supplying qualitative and quantitative security metrics. An important task in this second effort is to build an accountability system that will help monitor state. 
\paragraph{}
Finally, objects should be able to defend themselves against network failures and attacks. All protocols should incorporate mechanisms that respond to abnormal situations and allow the object to gracefully degrade its service. Objects should be able to use intrusion-detection systems and other defensive mechanisms to ward off attackers. Once an attack affects their services,
IoT elements should be able to act quickly to recover from any damage. Such elements can use feedback from other mechanisms and IoT entities to map the location of unsafe zones, where an attack has caused service outages.
\paragraph{}
The IoT is already more than a concept. By complying with security requirements, it can fully bloom into a paradigm that will improve many aspects of daily life. Open problems remain in many areas of the security view, such as cryptographic mechanisms, network protocols, data and identity management, user privacy, selfmanagement, and trusted architectures \cite{ref11}. 

\subsection{Identity management and communication issues} 
The diversity of identities, types and relationships configurations requires a judicious identity management, according to certain object identity principles shown in the following exemple: 
\begin{itemize}
\item[-] An object’s identity is not the same as the identity of its underlying mechanisms. The x-ray machine in the radiology department might have an IP address, but it should also have its own identity to distinguish it from other machines.  
\item[-]An object can have one core identity and several temporary identities. A hospital can become a meeting place for a health conference or a shelter after a fire.  
\item[-]An object can identify itself using its identity or its specific features. A virtual food identifies itself by its ingredients and quantity. 
\item[-]Objects know the identity of their owners. The device that controls a user's glucose level should know how that information fits in that user’s overall health.  
\end{itemize}
\emph{Identification: }The function of identification is to map a unique identifier or UID (globally unique or unique within a particular scope), to an entity so as to make it without ambiguity identifiable and retrievable. According to \cite{ref2}, the identification process could be done basically in two ways. The fist one is to physically tag one object by means of RFIDs (or similar). In such a way, an object can be read by means of an appropriate device, returning an identifier that can be looked up in a database for retrieving the set of features associating to it. The second possibility is to provide one object with its own description, it will be directly communicated through wireless connection. These two approches can complement each other. RFID-based identification is cheaper in term of the electronics to be embedded in the objects, but requires a database access where information about an object is stored. The self-description based approach, on the contrary, relaxes the requirements to access to a global database, but still requires to embed more electronics into everyday objects. 
\paragraph{}
\emph{Authentication: }Authentication is the prossess of proving identity,  it is an important part of identity management, objects will have to handle different behaviour according to it. A house could have several appliances that only certain residents and visitors can use at particular times. The refrigerator could lock  itself after midnight to any resident or visiting teenagers, but remain open for the adults. A promising approach is diverse authentication methods for humans and machines. A user could open an office door using bioidentification (such as a fingerprint) or an object  within a personal network, such as a passport, identity card, or smartphone. Combining authentication  methods can prevent any loss of overall system security. Such combinations typically take the form of what I am + what I know or what I have + what I know \cite{ref11}.
Because the IoT deals with multiple contexts, an entity is not likely to reveal its identity all the time. For example, in a vehicular network, a police car can reveal its identity to cars and staff at the police station, but keep its identity hidden during undercover work unless it is interacting with other police cars.  
 \paragraph{}
\emph{Authorization: }Authorization is also an identity management concern. Authentication and authorization share open  research issues, such as finding a balance between centralized and distributed systems to answer the question of who is in charge of defining and publishing roles. However, specific topics, such as delegation, fall mainly under the authorization umbrella. An IoT element can delegate certain tasks to other objects for a limited time. For example, an object in the user’s personal network, such as his phone, can check on his behalf to see if his suitcase contains all the clothes needed for an upcoming conference. The services, an object provides, might depend on the number of credentials presented. For example, a classroom could provide anyone who asks with the name of the course being taught, but it would release the syllabus of that course only to students with authorization certificates from the dean.
\paragraph{}
The futur deployement of IoT requires the developement of advanced techniques able to embed communication capabilities into every-day objects. In the last years, researches have been led on low-cost, low-power consumption and micro/nanpo-electronics.
Low-power communications is a well-established research field within the sensor networking community. The typical approach pursued in such works relates to the match of the RF front-end activation patterns (i.e., sleep periods) to the traffic pattern. The use of such protocols, however at present, does not provide a final answer to the optimization of energy consumption versus scalability issues.  These are of paramount importance for IoT scenarios, as battery replacement is a costly process to be avoided as much as possible, especially for large scale deployments. Furthermore, the basic idea of such protocols is to perform active/sleep duty cycles in order to save the power dispersed in idle listening.
More recently, advances in the field of nano-scale accumulators as well as energy harvesting techniques appear of prominent interest to limit the need for battery replacements \cite{ref17}.
\subsection{Ubiquitous intelligence}
The Internet of Things will create a dynamic network of billions or trillions of wireless identifiable (things) communicating with one another and integrating the developments from concepts like Pervasive Computing, Ubiquitous Computing and Ambient Intelligence. Internet of Things hosts the vision of ubiquitous computing and ambient intelligence enhancing them by requiring a full communication and a complete computing capability among things and integrating the elements of continuous communication,
identification and interaction. The Internet of Things fuses the digital world and the physical world by bringing different concepts and technical components together: pervasive networks, miniaturization of devices, mobile communication, and new models for business processes.
\paragraph{}
IoT scenarios will be typically characterized by huge amounts of data made available. A challenging task is to interpret such data and reason about it. This underpins the need to have an actionable representation of IoT data and data streams. This represents a key issue in order to achieve re-usability of components and services, together with interoperabilty among IoT solutions. Advances in data mining and knowledge representation/management will also be required, to satisfactorily address the peculiar features of IoT technologies. A related research field is that of distributed artificial intelligence, which addresses how autonomous software entities, usually referred to as (agents), can be made able to interact with the environment and among themselves in such a way to effectively pursue a given global goal\cite{ref15}.
\paragraph{}
IoT may well inherit concepts and lessons learned in pervasive computing, ambient intelligence applications and service-oriented computing. Researchers working in the field of human-computer interfaces and user-centric design methodologies, in particular, addressed already several issues concerning the impact of sensorized and pervasive environment on the user experience. Since IoT will take the reference scenarios one step further in terms of scale and offered features, it will also require the development of suitable, scalable service delivery platforms that permit multiple services to coexist.
\subsection{Standardisation efforts}
Although considerable efforts have been made to standardize the IoT paradigm by scientific communities, European Standards
Organizations (ETSI, CEN, CENELEC,etc.), Standardization Institutions (ISO, ITU) and global Interest Groups and Alliances (IETF, EPCglobal, etc.), they are not integrated in a comprehensive framework.
\paragraph{}
Efforts towards standardization have focused on several principal areas: RFID frequency, protocols of communication between readers and tags, and data formats placed on tags and labels. EPCglobal, European Commission and ISO are major standardization bodies dealing with RFID systems. EPCglobal mainly aims at supporting the global adoption of a EPC for each tag and related industry driven standards. European Commission has made coordinated efforts aiming at defining RFID technologies and supporting the transition from localized RFID applications to the IoT. Differently from these, ISO deals with how to modulate, utilize frequencies and prevent collision technically.
\paragraph{}
The European Telecommunications Standards Institute (ETSI) has launched the Machine-to-Machine (M2M) Technical Committee to conduct standardization activities relevant to M2M systems and define cost-effective solutions for M2M communications. Due to lack of standardization of this leading paradigm towards IoT, standard Internet, Cellular and Web technologies have been used for the solution of standards. Therefore, the ETSI M2M committee aims to develop and maintain an end-to-end architecture for M2M (with end-to-end IP philosophy behind it), and strengthen the standardization efforts on M2M. Within the Internet Engineering
Task Force (IETF), there are two working groups 6LoWPAN and ROLL dealing with integrating sensor nodes into IPv6 networks.
\paragraph{}
6LoWPAN is to define a set of protocols to make the IPv6 protocol compatible with low capacity devices. Core protocols have been already specified. While ROLL recently produced the RPL (pronounced "ripple") draft for routing over low power and lossy networks including 6LoWPAN. Lots of contributions are needed to reach a full solution \cite{ref7}.
\subsection{Social and political issues}
The Internet has long since changed from being a purely informational system to one that is socio-technological and has a social, creative and political dimension. But the importance of its non-technological aspects is becoming even more apparent in the development of Internet of Things, since it adds new challenges to these non-technological aspects.
\paragraph{}
Several critical questions need to be asked with regard to possible consequences of the full IoT's deployement, much of the public debate on whether to accept or reject the Internet of Things involves the conventional dualisms of “security versus freedom” and “comfort versus data privacy”. In this respect, the discussion is not very different from the notorious altercations concerning store cards, video surveillance and electronic passports. As with RFID, the unease focuses mainly on personal data that is automatically collected and that could be used by third parties without people’s agreement or knowledge
for unknown and potentially damaging purposes.
\paragraph{}
Personal privacy is indeed coming under pressure. Smart objects can accumulate a massive amount of data, simply to serve us in the best possible way. Since this typically takes place unobtrusively in the background, we can never be entirely sure whether we are being (observed) when transactions take place.
Aside of the data protection issues, there is also the question of who would own the masses of automatically captured and interpreted real-world data, which could be of significant commercial or social value, and who would be entitled to use it and within what ethical and legal framework.
\paragraph{}
Another critical aspect is that of dependence on technology. In business and also in society, generally, we have already become very dependent on the general availability of electricity; infrequent blackouts have fortunately not yet had any serious consequences. But if everyday objects only worked properly with an Internet connection in the future, this would lead to an even greater dependence on the underlying technology.
If the technology infrastructure failed for whatever reason: design faults, material defects, sabotage, overloading, natural disasters or crises, it could have a disastrous effect on the economy and society. Even a virus programmed by some high-spirited
teenagers that played global havoc with selected everyday objects and thus provoked a safety-critical, life-threatening or even politically explosive situation could have catastrophic consequences.
\paragraph{}
The Internet of Things has now arrived in politics. A study for the (Global Trends 2025) project carried out by the US National Intelligence Council states that foreign manufacturers could become both the single source and single point of failure for mission-critical Internet-enabled things, warning not only of the nation becoming critically dependent on them, but also highlighting the national security aspect of extending cyberwars into the real world: U.S. law enforcement and military
organizations could seek to monitor and control the assets of opponents, while opponents could seek to exploit the United States.
\paragraph{}
The European Commission is reflecting vocally but somewhat vaguely on the problem of governance for a future Internet of Things. The issue here is how to safeguard the general public interest and how to prevent excessively powerful centralized structures coming into being or the regulatory power of the Internet of Things falling exclusively into the hands of what they describe as a single specific authority \cite{ref14}.
\paragraph{}
To be truly beneficial, Internet of Things requires more than just everyday objects equipped with microelectronics that can cooperate with each other. Just as essential are secure, reliable infrastructures, appropriate economic and legal conditions and a social consensus on how the new technical opportunities should be used. This represents a substantial research issue for the futur.
\section{Conclusion}
Internet Of Things brings the possibility to connect billions of every-day's objects to the Internet, allowing them to interact and to share data. This prospect open new doors toward a futur where the real and virtual world merge seamlessly through the massive deployment of embedded devices. This survey has aimed to provide a brief overview of the IoT's state-of-art, including the main definitions and visions, enabling technologies, already or soon available applications and open research issues focusing on the security perspective. The IoT has the potential to add a new dimension to the ICT sector by enabling communications with and among smart objects, thus leading to the vision of (anytime, anywhere, anymedia, anything) communications. Though a lot still to be done in order to fullfil the IoT vision.
\bibliographystyle{plain} 
\bibliography{ref} 

\begin{thebibliography}{10}

\bibitem{ref4}
The epcglobal architecture framework.
\newblock page~19, March 2009.

\bibitem{ref2}
Francesco De Pellegrini Imrich~Chlamtac Daniele~Miorandi, Sabrina~Sicari.
\newblock Internet of things: Vision, applications and research challenges.
\newblock {\em Ad Hoc Networks}, pages 1497--1516, april 2012.

\bibitem{ref7}
Yi-Duo~Liang De-Li~Yang, Feng~Liu.
\newblock A survey of the internet of things.
\newblock {\em The 2010 International Conference on E-Business Intelligence},
  2010.

\bibitem{ref9}
R.~Wichert E.~Aarts.
\newblock Ambient intelligence.
\newblock {\em Technology Guide, Springer, Berlin, Heidelberg}, pages 244--249,
  2009.

\bibitem{ref17}
N.~Harris-B. Al-Hashimi G.~Merrett, N.~White.
\newblock Energy-aware simulation for wireless sensor networks.
\newblock {\em Proceedings of IEEE SECON, Rome, Italy}, pages 64--71, 2009.

\bibitem{ref15}
Peter Friess-Sylvie~Woelfflé Harald~Sundmaeker, Patrick~Guillemin.
\newblock Vision and challenges for realising the internet of things.
\newblock {\em Cluster of European Research Projects on the Internet of
  Things}, March 2010.

\bibitem{ref16}
Y.~Sankarasubramaniam-E.~Cayirci I.F.~Akyildiz, W.~Su.
\newblock Wireless sensor networks: a survey.
\newblock {\em Computer Networks 38 (4)}, pages 393--422, 2002.

\bibitem{ref8}
D.Culler J.~Hui and S.~Chakrabarti.
\newblock Incorporating ieee 802.15.4 into the ip architecture-internet
  protocol for smart objects.
\newblock {\em Ad Hoc Network}, jan 2009.

\bibitem{ref1}
Slaven Marusic-Marimuthu~Palaniswami Jayavardhana~Gubbi, Rajkumar~Buyya.
\newblock Internet of things (iot): A vision, architectural elements, and
  future directions.
\newblock {\em journal}, page~24, jan 2007.

\bibitem{ref3}
Giacomo~Morabito Luigi~Atzori, Antonio~Iera.
\newblock The internet of things: A survey.
\newblock {\em Computer Networks}, page~19, May 2010.

\bibitem{ref10}
F.~Rosenberg M.~Baldauf, S.~Dustdar.
\newblock A survey on context-aware systems.
\newblock {\em Int. J. Ad Hoc Ubiquitous Comput}, pages 263--277, april 2007.

\bibitem{ref6}
A.~Gluhak M.~Presser.
\newblock Eurescom message-the magazine for telecom insiders.
\newblock {\em Ad Hoc Networks}, feb 2009.

\bibitem{ref14}
Floerkemeier~C Mattern, F.
\newblock Vom internet der computer zum internet der dinge.
\newblock {\em Informatik-Spektrum 33(2)}, pages 107--121, 2010.

\bibitem{ref13}
J.~Lacy M.Blaze, J.~Feigenbaum.
\newblock Decentralized trust management.
\newblock {\em Proceedings of IEEE nternational Symposium Security and Privacy,
  Colorado Springs}, pages 164--173, 1996.

\bibitem{ref11}
Pablo~Najera Rodrigo~Roman and Javier Lopez.
\newblock Securing the internet of things.
\newblock {\em IEEE Computer}, pages 51--58, september 2011.

\bibitem{ref5}
K.~Sakamura.
\newblock Challenges in the age of ubiquitous computing: a case study of
  t-engine - an open development platform for embedded systems.
\newblock {\em Proc.Of the 28th International Conference on Software
  Engineering}, pages 713--720, May 2006.

\bibitem{ref12}
Xiaohui~Lang Xu~Li, Rongxing~Lu and Xuemin~(Sherman) Shen.
\newblock Smart community: An internet of things application.
\newblock {\em IEEE Communicatins Magazine}, pages 68--75, november 2011.

\end{thebibliography}
\end{document}